\documentclass[aip, amsmath, amssymb, reprint]{revtex4-1}

\usepackage{graphicx}
\usepackage{dcolumn}
\usepackage[T1]{fontenc}
\usepackage{mathptmx}
\usepackage{times}
\usepackage{amsmath}
\usepackage{bm}
\usepackage{subfigure}
\usepackage[utf8]{inputenc}
\usepackage{hyperref}
\usepackage{xcolor}

\newcommand{\e}{\varepsilon}
\newcommand{\f}{\varphi}

\renewcommand{\Re}{{\rm Re}}
\renewcommand{\Im}{{\rm Im}}

\begin{document}
\preprint{AIP/123-QED}

\title[]{Maximal Interaction of Electromagnetic Radiation with Corona-Virions}

\author{Constantinos Valagiannopoulos}
 \email{konstantinos.valagiannopoulos@nu.edu.kz}
\affiliation{School of Sciences and Humanities, Nazarbayev University, KZ-010000, Kazakhstan}
\author{Ari Sihvola}
 \email{ari.sihvola@aalto.fi}
\affiliation{School of Electrical Engineering, Aalto University, FI-00076, Finland}

\date{\today}

\begin{abstract}
Absorption and scattering of the impinging electromagnetic waves are the two fundamental operations describing the energy exchange of any, organic or inorganic, particle with its environment. In the case of virion cells, substantial extinction power, counting both absorbing and scattering effects, is a prerequisite for performing a variety of coupling actions against the viral particles and, thus, a highly sought-after feature. By considering realistic dispersion for the dielectric permittivity of proteins and a core-shell modeling allowing for rigorous formulation via Mie theory, we report optical extinction resonances for corona-virions at mid-infrared range that are not significantly perturbed by changes in the objects size or the background host. Our findings indicate the optimal regime for interaction of photonic radiation with viral particles and may assist towards the development of equipment for thermal damage, disintegration or neutralization of coronavirus cells.
\end{abstract}

\maketitle

\section{\label{intro} Introductory Comment}
The energy exchange between an impinging light beam and virus particles is a cornerstone in several biophotonic operations with important applications in Microbiology, Pharmacology, Medical Physics and Biochemistry. The study of this interaction between virions and electromagnetic fields has been greatly assisted by a large set of analytical methods and physical concepts reported for inorganic particles \cite{BohrenBook}. That inevitably led in their implementation and translation to understand how microbes absorb and re-emit visible light in various directions \cite{MicrobialLaser} or what are the biophysical processes behind ultraviolet germicidal irradiation \cite{KowalskiHandbook}. Furthermore, applications of diverse types of scattering techniques to systems of microorganisms have been reviewed \cite{ApplicationsLightScat} while the diffusion behavior of viral macromolecules into liquids has been extensively elaborated \cite{DynamicLightScat}. Importantly, highly sensitive virus detection has become possible based on optical trapping \cite{OpticalTrap} or via the reactive shift of a whispering-gallery mode \cite{SingleVirusDetection}, facilitating convenient medical diagnosis and food inspection. In addition, the coupling of the incident beams with virions has been utilized in measuring the refractive index of the cells with high precision \cite{AnimalVirus} and analyzing single viruses with a resolution comparable to that of electron microscopy \cite{LabelFreeAnalysis}.

Mie theory \cite{HulstBook}, admitting rigorous solution to the scattering of electromagnetic waves by multilayered spheres, makes a powerful tool for treating similar problems involving radiation impinging on virus particles, due to their quasi-spherical shapes. In particular, simple formulas are derived towards the interpretation of the characteristic anomalies in the optical activity of membrane suspensions \cite{OpticalActivity} of for the evaluation of collective backscattering from abundant viruses into sea water \cite{ViralSuspensions}. Mie scattering has been also used to model the light intensity produced by virion-like nano-objects in biosensors \cite{VirusLikeParticles}, flow cytometers \cite{CharacterizingLight} and phase microscopes setups \cite{NuclearRefIndex}. Finally, analytical expressions have been provided for the shifts in the resonance frequencies of spherical dielectric microresonators owing to plasmonic nanoparticles \cite{TheoryResonanceShifts} or protein binding \cite{UltrasensitiveDetection}, paving the way to highly-efficient bioimaging.

\begin{figure}[ht!]
\centering
\subfigure[]{\includegraphics[height=4.2cm]{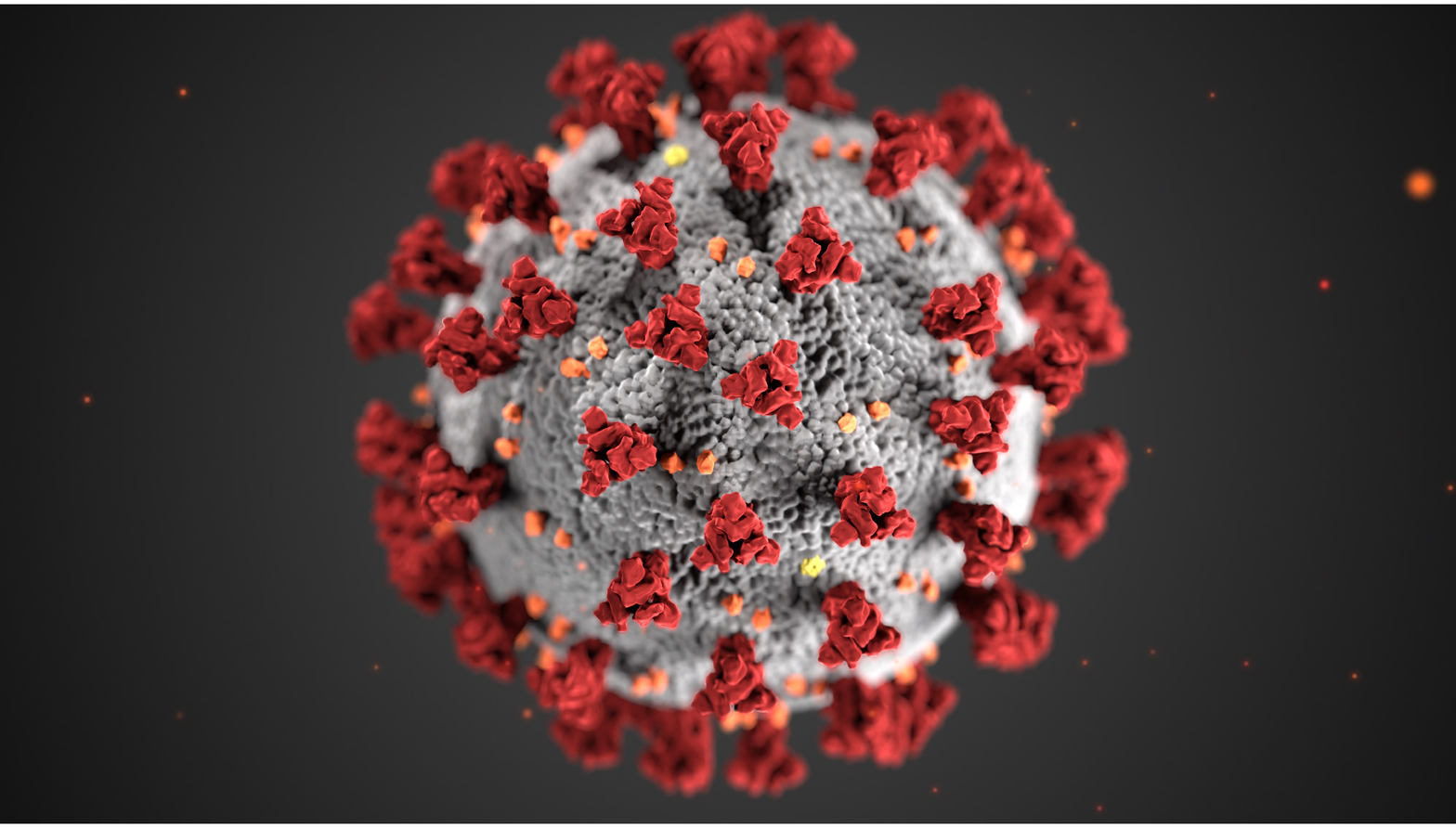}
   \label{fig:Fig1a}}
\subfigure[]{\includegraphics[height=4.2cm]{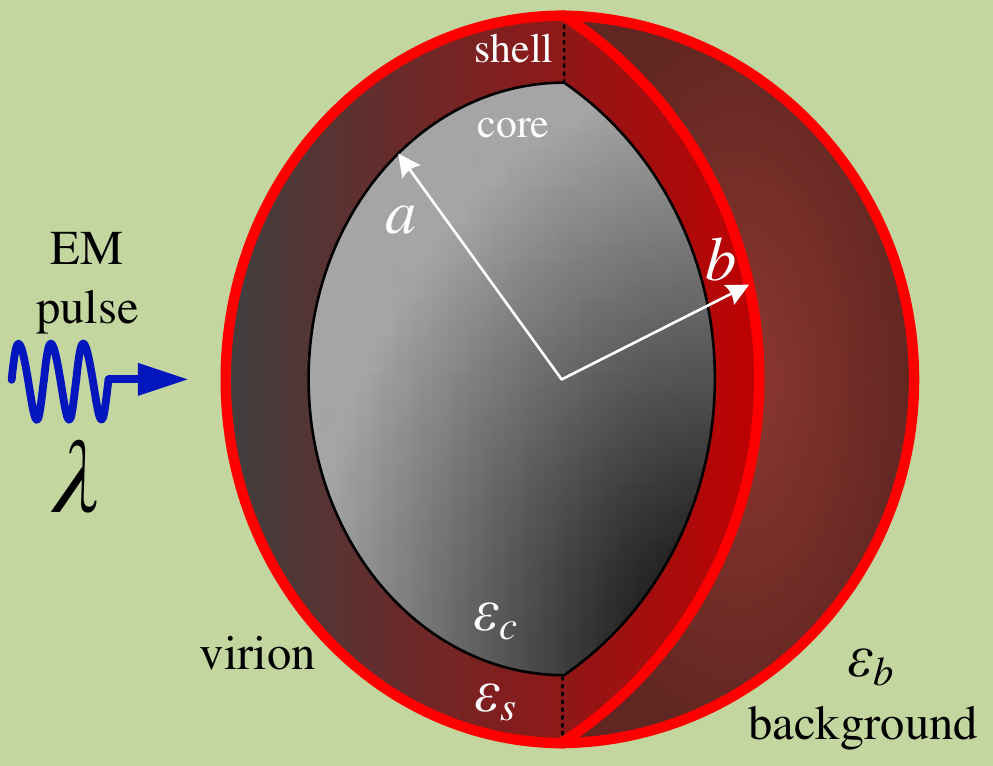}
   \label{fig:Fig1b}}
\caption{(a) Illustrative representation revealing the coronavirus morphology when viewed via electron microscope. Both the gray core and the brown spikes, that adorn the outer surface of the virus imparting the look of a corona surrounding the virion, are made of proteins. The image has taken from Public Health Image Library (\href{https://phil.cdc.gov/Details.aspx?pid=23311}{PHIL}) and is free of any copyright restrictions. (b) The adopted model in the present study; the spikes are replaced by an isotropic shell of permittivity $\e_s$ and thickness $(b-a)$ surrounding a homogeneous core of permittivity $\e_c$ and radius $a$; the structure is hosted by a background medium of permittivity $\e_b$. The virion is bombarded by an electromagnetic (EM) pulse of high intensity and central wavelength $\lambda$.}
\label{fig:Figs1}
\end{figure}

Coronaviruses constitute a special category of viruses whose genome is hosted into protein cells, with rod-shaped spikes projecting from their surfaces; these elongated bumps, when seen though electron microscope, create an image mimicking the solar corona, to which the viruses owe their name. Since the 2020 global pandemic outbreak \cite{ANewCoronavirus, NovelCoronavirus}, the whole world became familiar with the term ``coronavirus'' while medical scientists struggle to handle \cite{StrategicApproach} that continued threat, responsible for hundreds of thousands of deaths and unprecedented socio-economic damage. Due to the alarming situation, numerous experimental efforts have been devoted to test the photonic response of that specific coronavirus (SARS-CoV-2) for various objectives like the fast biosensing that secures reliable viral disease diagnosis \cite{DualFunctional} or the development of cellular nanosponges that are allegedly able to neutralize the virus \cite{CellularNanosponge}. Interestingly, several studies also indicate that phototherapy has immense potential to reduce the impact of coronavirus diseases \cite{LightAsTreatment}, and offers suggested ways that the healthcare industry can integrate modern light technologies in the fight against SARS-CoV-2 and its mutant versions.

However, analytical modeling via Mie theory is very rarely \cite{PhotopolarimetricalProperties} involved to works studying the light-coronavirus interactions, despite its rigorous yet simple formulation. In this paper, we systematically examine the Mie scattering of a protein nanosphere covered by a suitably homogenized shell emulating the presence of amine radial spikes. The size of the core \cite{ArchitectureSelfAssembly} and the spikes length \cite{SpikeProteins} are varying between their realistic limits, according to the adopted conceptual layouts \cite{SpikeProteins} and the images taken from electron microscopes \cite{FusionCore}. As far as the background media that host the virions are concerned, they are also selected based on the presently available data, since the virus can transmit through the air \cite{AirHost}, while can exist into blood \cite{BloodHost} and the human organs like liver \cite{LiverHost} as well. The impinging electromagnetic pulse is harmonic with frequency that spans from the hard ultraviolet to the long infrared part of the spectrum; the dispersion of the incorporated materials across this extensive band is taken into account, based on experimental measurements contained in well-established references \cite{ProteinData, BloodData, LiverData}. Our aim is to maximize the extinction power from the core-shell nanoparticle, which is a prerequisite for any action against the virion from disintegration to isolation. The influence of the geometric characteristics of the considered model on the observable quantity is identified and the optimal mid-infrared wavelengths leading to substantial extinction are determined. Importantly, the reported resonances are found highly insensitive to structural changes regardless of the background and thus are applicable to ensembles of corona-virions of diverse features. Our findings may inspire clinical research towards the development of diagnostic products and devices that require significant power interaction with SARS-CoV-2 particles to dissolve or neutralize them.

\section{\label{appr} Proposed Approach}

\subsection{Core-Shell Model}
As shown in the illustrative sketch of Fig. \ref{fig:Fig1a}, a typical corona-virion is composed of a homogeneous core (nucleocapsid), containing a mixture of proteins \cite{ArchitectureSelfAssembly}, surrounded by radial spikes of glycoprotein \cite{StructureFunction}. We treat all proteins as diisopropylamine, also known as DIPA, whose dispersive permittivity $\e_c=\e_c(\lambda)$ can be easily found. As far as the crown of protein rods is concerned, the simplest way to model it is as a homogeneous shell of permittivity given by a weighed sum of the protein dielectric constant $\e_c$ and that of the background medium $\e_b$:     
\begin{eqnarray}
\e_s=(1-s)\e_b+s \e_c,
\label{ShellPermittivity}
\end{eqnarray}
where $0<s<1$ is the filling factor indicating the percentage of the corona volume occupied by the spikes. 

We could follow alternative and more accurate approaches to model the photonic setup of the virion, like approximating the shell by quasi-homogeneous multilayers \cite{QuasiHomog}, considering radial anisotropy \cite{HomogMulti, SihvolaSphereA} or even assuming systropic properties for the fabric of the spherical particles \cite{SihvolaSphereB}. In this case, the scalar $\e_s$ would have been replaced by a uniaxial tensor $[\e_s]={\rm diag}(\e_r,\e_t,\e_t)$ in spherical coordinates $(r,\theta,\f)$. The radial permittivity $\e_r=\e_s$ will be given by \eqref{ShellPermittivity} since the corresponding depolarization factor of a needle vanishes. On the contrary, the transversal constant (along local $\theta,\f$ directions), $\e_t$, will be written as \cite{SihvolaMixingForm}: 
\begin{eqnarray}
\e_t=\e_b\frac{(\e_b+\e_c)-s(\e_b-\e_c)}{(\e_b+\e_c)+s(\e_b-\e_c)}.
\label{TransversalPermittivity}
\end{eqnarray}
The solution of the wave equation into such a medium involves cylindrical Bessel functions $J_{\nu},Y_{\nu}$ of orders \cite{ElectromagneticTransparency}:
\begin{eqnarray}
\nu=\nu_n=\frac{1}{2}\sqrt{1+4\frac{\e_t}{\e_r}n(n+1)},
\label{BesselOrders}
\end{eqnarray}
for $n\in\mathbb{N}^*$.

However, the size of the virion is small (average radius $b\cong 60$ nm) and the regarded wavelengths large (average $\lambda\cong 3500$ nm, mid-infrared); accordingly, the incoming beams are not expected to ``feel'' a more advanced setup emphasizing on the geometry details. In other words, a very slowly moving wave will perceive the very complicated actual structure of the protein spikes in the same way as a homogeneous cladding of texture determined by \eqref{ShellPermittivity}. Thus, we advocate that the model in \ref{fig:Fig1b}, where the core (permittivity $\e_c$) of radius $a$ is engulfed by the shell (permittivity $\e_s$) of size $(b-a)$ and host into a homogeneous background (permittivity $\e_b$), captures sufficiently the electromagnetic interactions and the underlying photonic power interplay between the incident fields and the particle. Note finally that the Bessel order \eqref{BesselOrders} becomes complex if the permittivities \eqref{ShellPermittivity},\eqref{TransversalPermittivity} have non-zero imaginary parts, which is the case in our consideration; as a result, numerical issues \cite{WatsonTransform} may emerge if the shell is assumed anisotropic. That makes an extra reason to follow the isotropic and homogeneous modeling via \eqref{ShellPermittivity}. 

\subsection{Mie-Theory Formulation}
We assume that the virion of Fig. \ref{fig:Fig1b} is illuminated by an electromagnetic beam in the form of monochromatic plane wave (with free-space oscillating wavelength $\lambda\equiv 2\pi/k_0$), traveling into the background host. As mentioned above, the symbols $(r,\theta,\f)$ are used for the related spherical coordinates centralized at the cell, while the equivalent Cartesian ones read $(x,y,z)$; the suppressed harmonic time is of the form $e^{+i 2 \pi c t/\lambda}$, where $c$ is the speed of light into free space. For simplicity and without loss of generality (due to the spherical symmetry), we assume that the incident wave propagates along $+z$ axis and its electric field vector is always parallel to $x$ axis oscillating with amplitude $E_0>0$ (measured in $Volt/meter$). This background field can be decoupled in two terms, each of which satisfies Maxwell's laws: one term with no radial electric component (TE) and another one with no radial magnetic component (TM). These terms can be expressed as series of spherical harmonics, which dictate the $\theta-$ and $\f-$ dependence of the field quantities in all the regions defined by the concentric and entire surfaces, according to the rigorous Mie theory \cite{HulstBook}. 

After imposing the necessary boundary conditions, the scattered fields for $r>b$ (electric field vector of TE set and magnetic one for TM set), are written as \cite{NANOSPHERE}:
\begin{eqnarray}
\textbf{E}_{scat}^{TE}=E_0 \sum_{n=1}^{+\infty}i^{-n}S_n^{TE}h_n(k_br) 
\left\{\begin{array}{c}-\hat{\bm{\theta}} \csc\theta p_n(\theta)\cos\f \\
     +\hat{\bm{\f}}p_n'(\theta)\sin\f \end{array}\right\}  \nonumber,
\label{ETEscat}\\
\textbf{H}_{scat}^{TM}=\frac{E_0\sqrt{\e_b}}{\eta_0} \sum_{n=1}^{+\infty}i^{-n}S_n^{TM}h_n(k_br)
\left\{\begin{array}{c}-\hat{\bm{\theta}} \csc\theta p_n(\theta)\sin\f \\
     -\hat{\bm{\f}}p_n'(\theta)\cos\f \end{array}\right\}  \nonumber,
\label{HTMscat}
\end{eqnarray}
where $p_n(\theta)=P_n^1(\cos\theta)$ is the Legendre polynomial of first order, degree $n$ and argument $\cos\theta$; in addition, $h_n$ is the spherical Hankel function of order $n$ and second type. The symbol $k_b=k_0\sqrt{\e_b}=2\pi\sqrt{\e_b}/\lambda$ stands for the wavenumber into the background medium and $\eta_0=120\pi~{\rm \Omega}$ the wave impedance into free space. The coefficients $S_n^{TE/TM},$ are complex dimensionless quantities and not shown here for brevity \cite{AchievingTransparency, QUANTUM}.

The power $P_{scat}$ carried by the TE and TM scattered component, which constitutes a self-consistent electromagnetic field into the background host, expresses how much the sphere perturbs the background field distribution externally to it. It can be easily computed with use of Poynting's theorem and expansions of $h_n(k_br)$ for large arguments $k_br\gg 1$ (in the far region), as follows \cite{SingleSeries}:
\begin{eqnarray}
P_{scat}=P_0\sum_{n=1}^{+\infty}\frac{n^2(n+1)^2}{2n+1}\left(\left|S_n^{TE}\right|^2+\left|S_n^{TM}\right|^2\right),
\label{Pscat}
\end{eqnarray}
where $P_0=\frac{\pi E_0^2}{k_0^2 \eta_0\sqrt{\e_b}}>0$ is a quantity measured in $Watt$ and $k_0$ is the free-space wavenumber. The power absorbed by the particle, given the presence of lossy constituent media, is evaluated by applying again Poynting's theorem but for the total field this time. Indeed, if we integrate the power spatial density across any sphere of radius $r>b$ (even the infinite $k_b r \rightarrow +\infty$ one), we obtain:  
\begin{eqnarray}
P_{abs}=-P_{scat}-P_0\sum_{n=1}^{+\infty}n(n+1)\Re\left[S_n^{TE}+S_n^{TM}\right].
\label{Pabs}
\end{eqnarray}
Obviously, $P_{scat}, P_{abs}>0$, which means that the series in (\ref{Pabs}) should converge to negative values smaller than $(-P_{scat}/P_0)$. In the absence of any losses, we have $P_{abs}=0$ and the aforementioned sum equals to $(-P_{scat}/P_0)$. 

\begin{figure}[ht!]
\centering
\subfigure[]{\includegraphics[height=4.2cm]{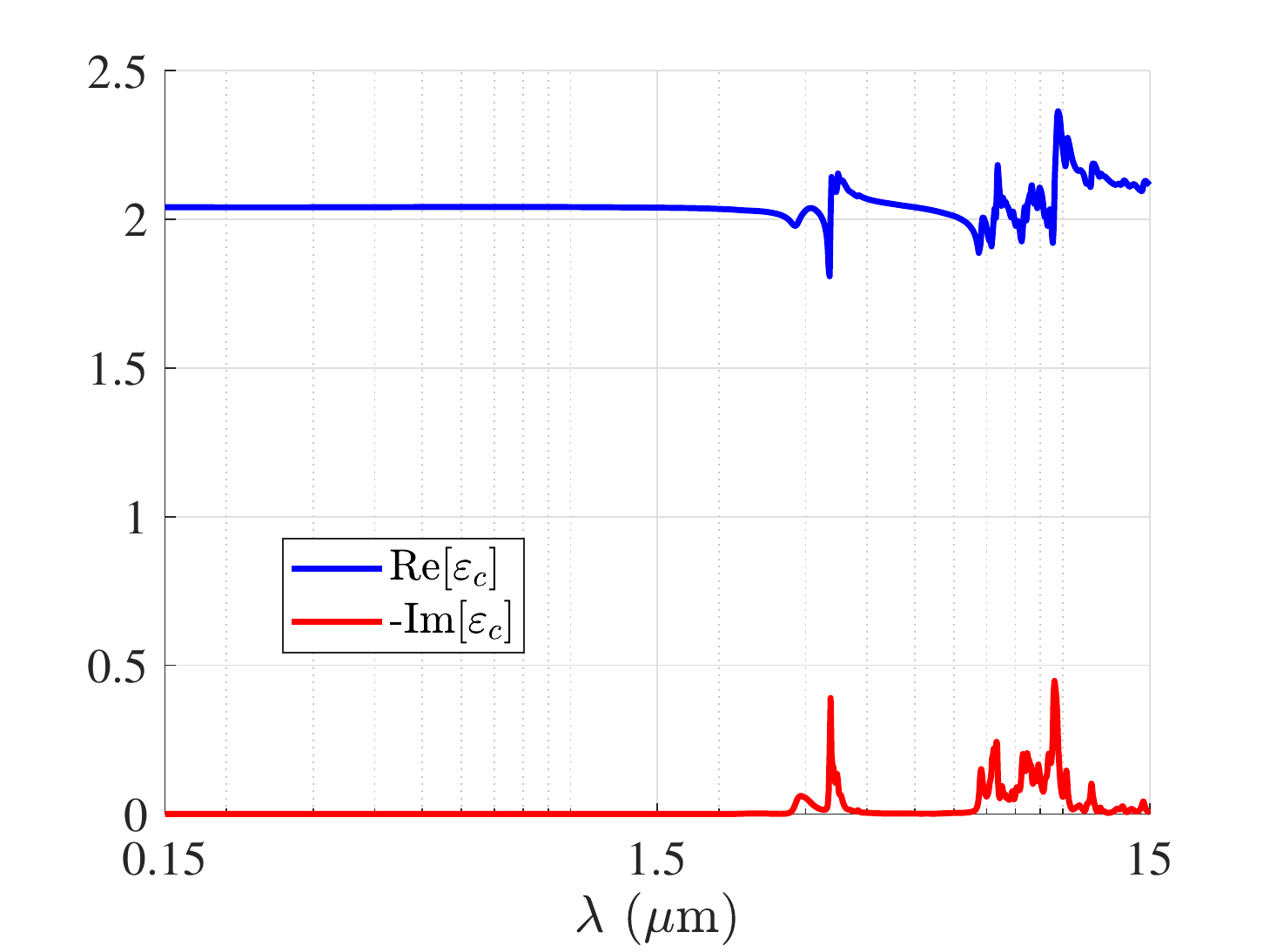}
   \label{fig:Fig12a}}
\subfigure[]{\includegraphics[height=4.2cm]{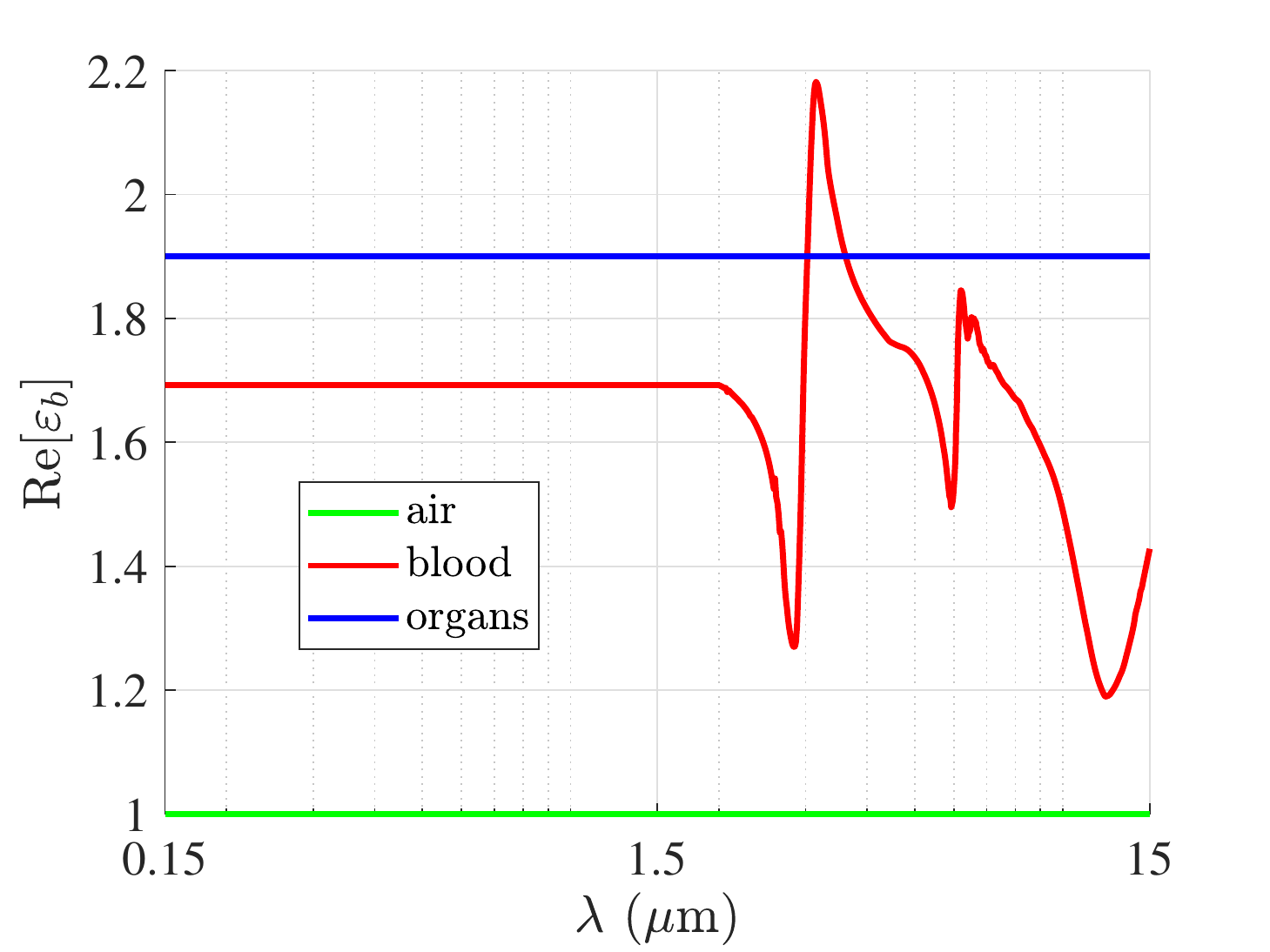}
   \label{fig:Fig12b}}
\caption{Dispersive permittivities $\e$ as functions of the incident wavelength $\lambda$. (a) Real and imaginary parts of the permittivity $\e_c$ for protein (diisopropylamine, DIPA). The data is obtained from well-established source \cite{ProteinData} and expanded to short wavelength limit, based on lossless assumption \cite{ProteinShortWavelength}.  (b) Real parts of the background permittivity $\e_b$ for alternative hosts: air, human blood and human organs (like liver). The various backgrounds are assumed to be lossless and the corresponding data has been obtained from reliable experimental measurements \cite{LiverData, BloodData}.}
\label{fig:Figs12}
\end{figure}

\subsection{Parameters and Observables}
Before proceeding to the numerical results and the discussion, let us first clarify the value ranges of the parameters incorporated into our model. In particular, the incident electromagnetic pulse is taken with a free-space central wavelength $\lambda$ belonging to extensive band spanning from hard ultraviolet short waves ($\lambda=150$ nm, UV-C) to long infrared radiation ($\lambda=15$ $\mu$m, IR-C). In addition, we consider an external radius for the virion varying into the interval: $30~{\rm nm}<b<100~{\rm nm}$, representing an assortment of various sizes \cite{NovelCoronavirus} and a radii ratio with $0.5<a/b<0.9$, corresponding to different lengths of the protein spikes \cite{SpikeInfo}. When it comes to the density $s$ of the spikes, we regard all possible values: $0<s<1$ from an absent crown ($s=0$) to a big homogeneous protein sphere of radius $b$ ($s=1$). 

The variation for the dispersive permittivity for the homogeneous core $\e_c=\e_c(\lambda)$ is depicted in Fig. \ref{fig:Fig12a}, where the data has been taken from reliable source \cite{ProteinData} and expanded to short wavelength limit \cite{ProteinShortWavelength}. We notice that the regarded protein (diisopropylamine, DIPA) is lossless across large parts of the wavelength spectrum except for two bands around $\lambda\cong 3.5~\mu{\rm m}$ and $\lambda\cong~9.5~\mu{\rm m}$; these losses are responsible for the corresponding variations of the real part $\Re[\e_c]$, according to the Kramers–Kronig relations that demand causal responses \cite{BohrenBook}. As far as the permittivity of the host is concerned, it is lossless and non-dispersive for air and human organs like liver \cite{LiverData} as shown in Fig. \ref{fig:Fig12b}. In the case of blood \cite{BloodData}, $\e_b$ exhibits some variation accompanied by moderate losses that are ignored ($\Im[\e_b]=0$) for a better formulation of the primary plane-wave excitation; otherwise a modification should be performed \cite{AbsorbingMedium}. Note that the real parts of the dielectric constants for the proteins, the human organs and the human blood are very close each other making a configuration of low textural contrast, where photonic power concentration is particularly challenging. Such a feature ``pushed'' us towards large operational frequencies (ultra-small wavelengths $\lambda$), as indicated above; only then the incoming light can interact strongly or resonate \cite{GeneralScattering} with an inclusion constituting a so mild perturbation of the refractive index.

When it comes to the output of the system, we normalize the absorbed and scattered power by the power of the incident illumination passing through the geometrical cross section of the scatterer, namely $P_{inc}=\frac{\pi E_0^2 b^2\sqrt{\e_b}}{2 \eta_0}$; in this way, the response of the virion becomes more meaningful. The most important quantity, however, is the extinction power $P_{ext}\equiv P_{scat}+P_{abs}$, namely the sum of scattering and absorption from \eqref{Pscat},\eqref{Pabs}, so it represents the total effect of the particle on radiation traveling into the background medium; thus, the basic metric of our study is given by:
\begin{eqnarray}
\frac{P_{ext}}{P_{inc}}=-\frac{2}{(k_0b)^2\e_b}\sum_{n=1}^{+\infty}n(n+1)\Re\left[S_n^{TE}+S_n^{TM}\right].
\label{MyMetric}
\end{eqnarray}
The series \eqref{MyMetric} is rapidly converging once the optical size $a/\lambda$ of the virion is small \cite{WatsonTransform}, which is the case in the present formulation. Our aim will be to identify the conditions under which that ratio $P_{ext}/P_{inc}$ is maximized so that the external electromagnetic beam couples optimally with the virion admitting its disintegration, conversion or neutralization.

\begin{figure}[ht!]
\centering
\subfigure[]{\includegraphics[height=4.2cm]{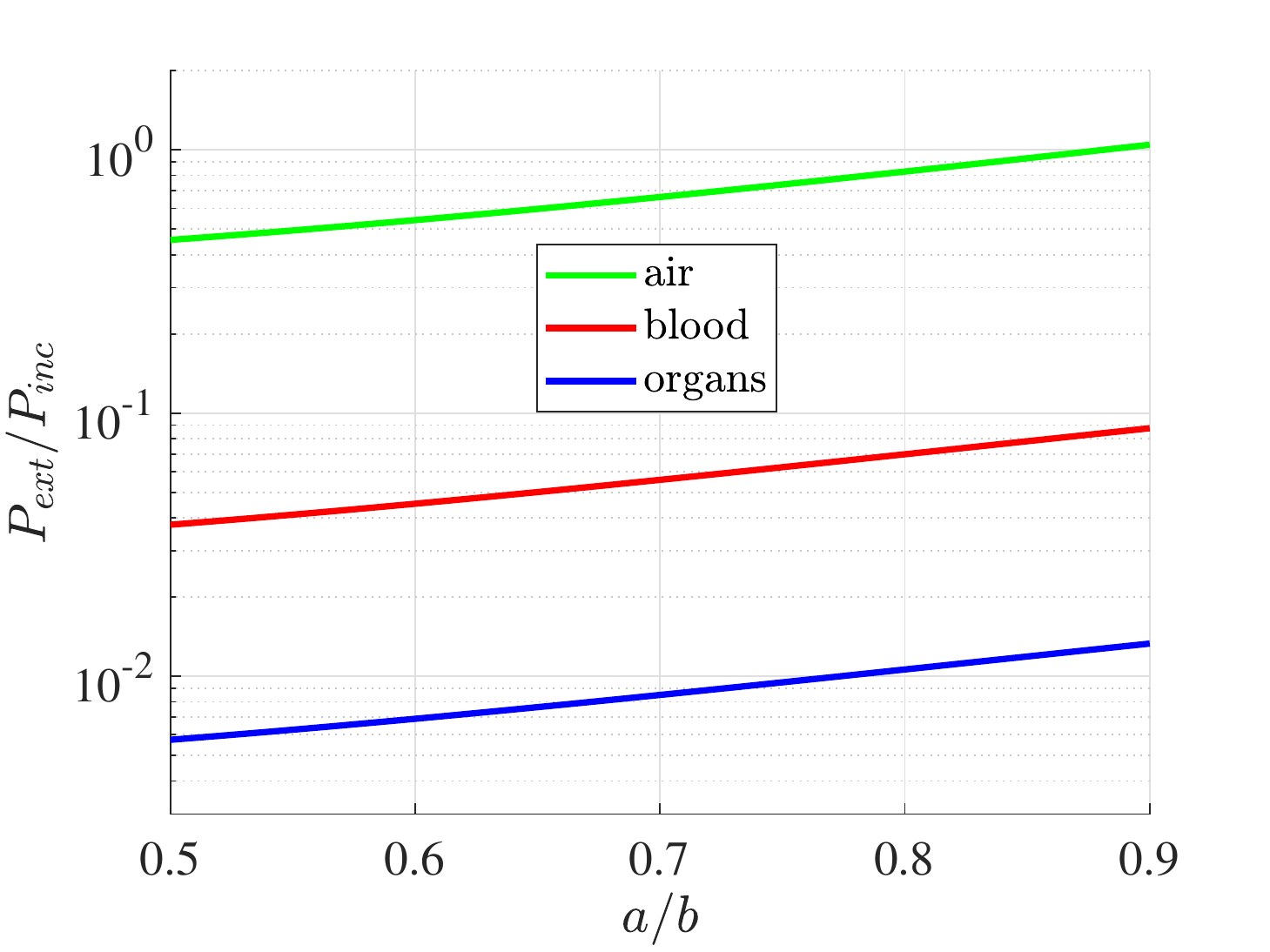}
   \label{fig:Fig2a}}
\subfigure[]{\includegraphics[height=4.2cm]{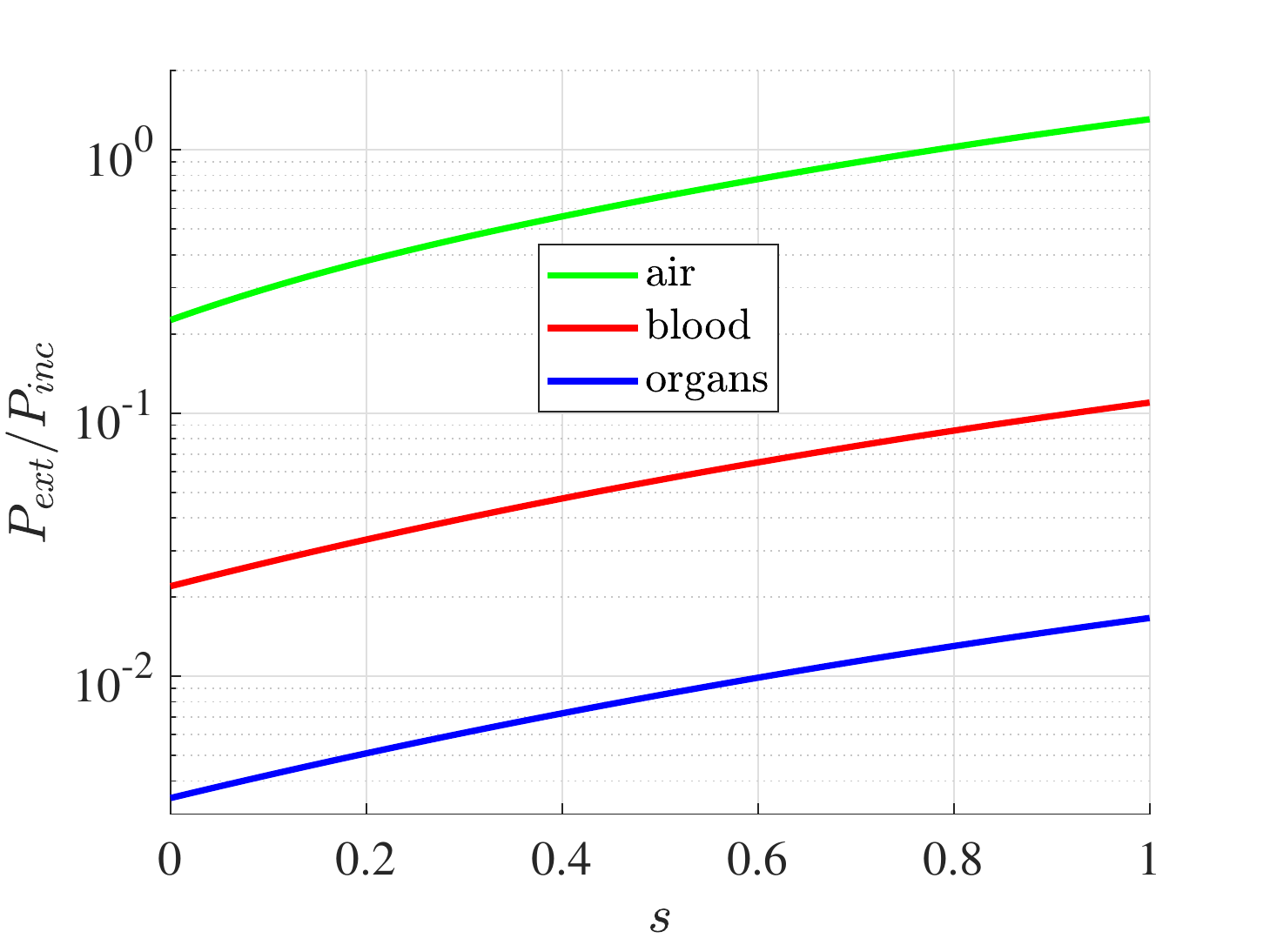}
   \label{fig:Fig2b}}
\caption{The extinction power $P_{ext}$ by the virion normalized by the incident power through the spherical cross section $P_{inc}$ for various backgrounds $\e_b$, as a function of: (a) the core-shell radii ratio $a/b$ ($s=0.5$) and (b) the filling factor $s$ of the shell ($a/b=0.7$). Plot parameters: $\lambda=250$ nm, $b=80$ nm. The effect of the material contrast is identified.}
\label{fig:Figs2}
\end{figure}

\section{\label{resdis} Results and Discussion}

\subsection{Maximal Extinction Power}
In Fig. \ref{fig:Fig2a}, we represent the extinction power $P_{ext}$ normalized by the incident power $P_{inc}$ with respect to the core-shell radii ratio $a/b$ for the three different background media indicated by Fig. \ref{fig:Fig12b}. We notice that the quantity showing how much the particle absorbs the radiation and ``shakes'' the local field distribution gets increased once the virion's corona becomes thinner indicating a more substantial power concentration around the core. In addition, the beneficial influence of the material contrast between the spherical cell and the background on $P_{ext}/P_{inc}$ can be identified since the highest values are recorded for airborne particles; on the contrary, the extinction power is small when $\e_b\cong\e_c$, as happening in the case of virions hosted into human organs. Quantitatively speaking, the magnitude $P_{ext}/P_{inc}$, at least for the adopted short wavelength ($\lambda\cong 250$ nm), is quite high and, for a densely populated crown by protein spikes, it surpasses unity. In other words, the presence of particle participates in a huge power exchange with its environment concerning the whole electromagnetic radiation passing through its transection. Obviously a $P_{ext}/P_{inc}>1$ is feasible since the object may interact with rays that are not directly incident on its spherical surface but travel in the vicinity of it \cite{MAXABS}.     

In Fig. \ref{fig:Fig2b}, we show the metric $P_{ext}/P_{inc}$ from \eqref{MyMetric} as a function of the filling factor $s$ for alternative hosts. A rapidly increasing trend is observed, demonstrating once more the amplifying effect of the needle-shaped rods on the particle-beam interaction. Furthermore, the textural contrast is again recognized as a factor that boosts the extinction power, while the measured quantity is somehow higher compared to Fig. \ref{fig:Fig2a}. It is also noticed that, despite the low difference between the refractive index of blood and organs as shown in Fig. \ref{fig:Fig12b}, the corresponding extinction power in human blood background is much higher; indeed, what counts is the spread between $\e_b$ and the protein's permittivity $\e_c$ as depicted in Fig. \ref{fig:Fig12a}.   

\begin{figure}[ht!]
\centering
\subfigure[]{\includegraphics[height=4.8cm]{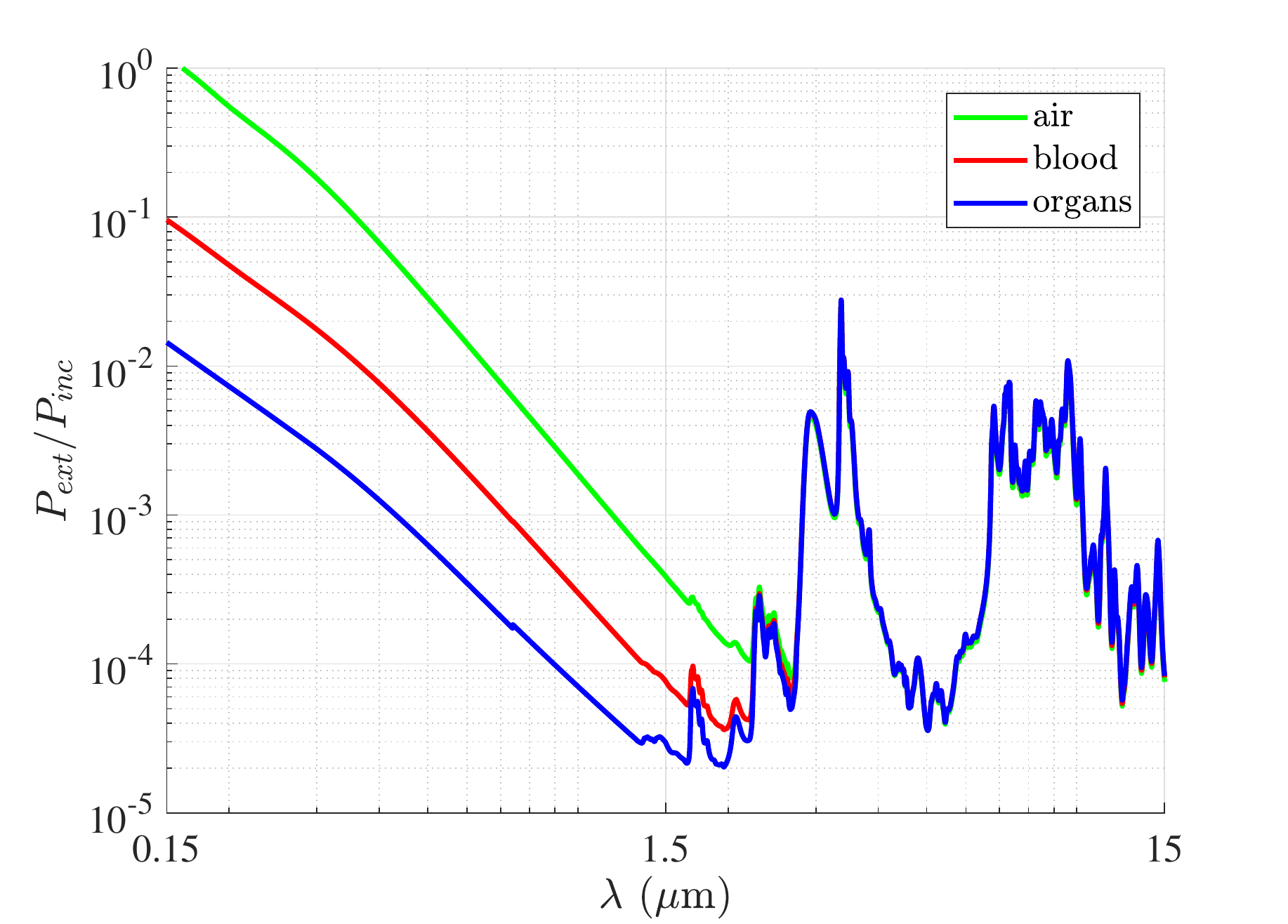}
   \label{fig:Fig3a}}
\subfigure[]{\includegraphics[height=4.8cm]{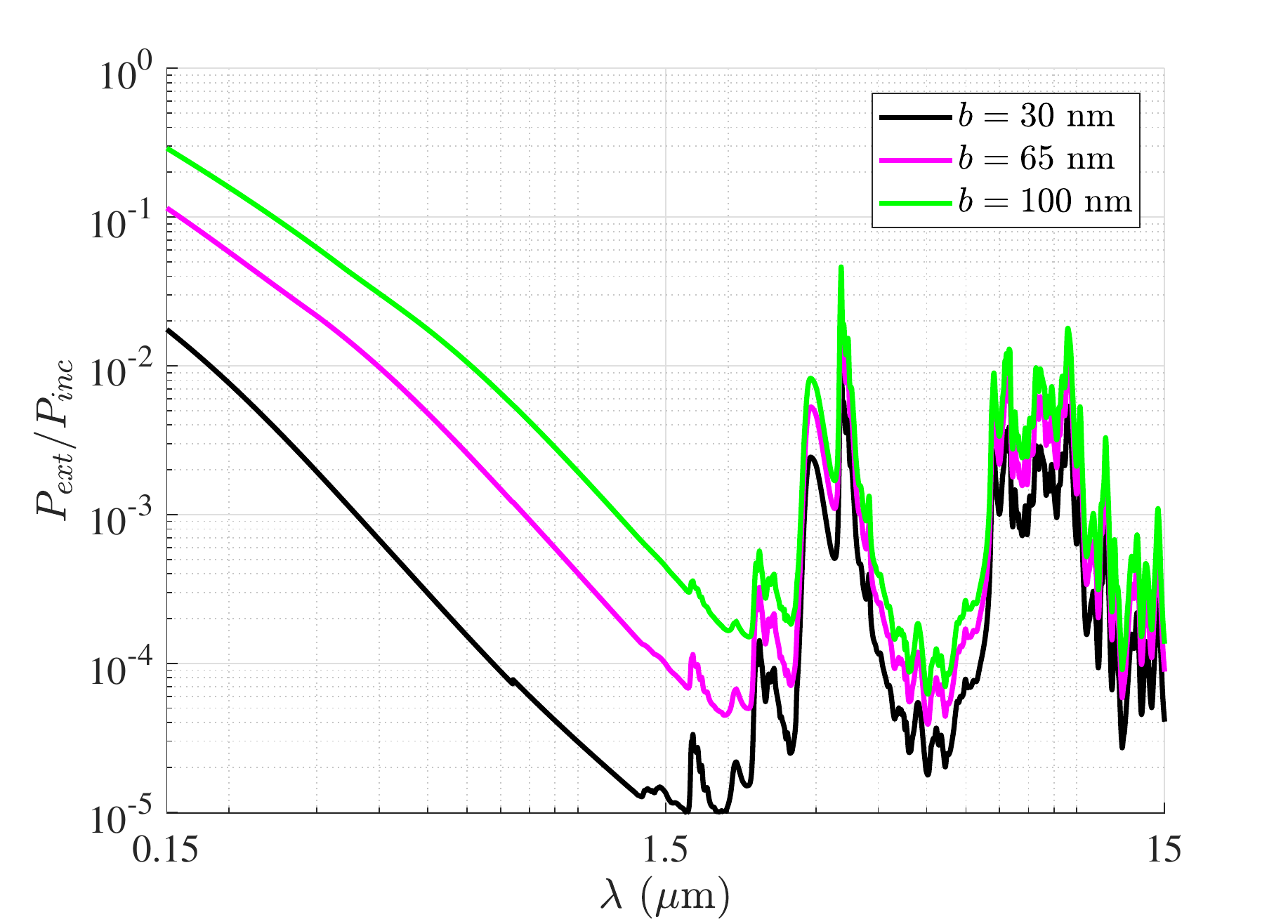}
   \label{fig:Fig3b}}
\caption{The extinction power $P_{ext}$ by the virion normalized by the incident power through the spherical cross section $P_{inc}$ as a function of the incoming wavelength $\lambda$ for: (a) various host environments $\e_b$ ($b=60$ nm), (b) various external sizes $b$ of the virions (in blood). Plot parameters: $a/b=0.7$, $s=0.5$. The strong local optimum at mid-infrared ($\lambda\cong 3370$ nm) is spotted, regardless of background and size.}
\label{fig:Figs3}
\end{figure}

Given the fact that Fig. \ref{fig:Figs2} concerns the interplay of the particle with the incoming illumination at ultra-violet frequencies, it is important to understand the response of the spherical virion across the whole of the considered band. In particular, in Fig. \ref{fig:Fig3a}, we represent the variation of the ratio $P_{ext}/P_{inc}$ with respect to oscillation wavelength $\lambda$ for the three regarded hosts. When the wavelengths are tiny ($\lambda<1$ $\mu$m), it is natural to spot a declining trend since the particle is not optically big enough to interact substantially with the incident electromagnetic pulse. However, beyond 1.5 $\mu$m, where proteins exhibit significant losses (see Fig. \ref{fig:Fig12a}), the metric increases by orders of magnitudes to reach a strong local maximum at $\lambda\cong 3.37$ $\mu$m, regardless of the background. Therefore, in order to maximally engage with that specific type of corona-virions, one should concentrate the impinging power in the spectral vicinity of that frequency; this conclusion is one of the major findings of our study. Similar resonances are exploited for protein molecule monolayers detection by covering the sensor to produce a controllable amount of resonance redshift \cite{ProteinDetection}. It should be finally stressed that the results are host-indifferent only when the operating wavelengths  $\lambda$ are large enough; for $\lambda<1$ $\mu$m, the relative order of $P_{ext}/P_{inc}$ for different hosts is dictated by the material contrast of Fig. \ref{fig:Figs12}, as happening in Fig. \ref{fig:Figs2}.  

In Fig. \ref{fig:Fig3b}, we repeat the calculations of Fig. \ref{fig:Fig3a} but for different sizes of virions $b$, all existing into human blood. The variation of $P_{ext}/P_{inc}$ is similar to that of Fig. \ref{fig:Fig3a} but the larger size makes a difference and increases, on average, the represented metric, even at lower frequencies ($\lambda>1.5$ $\mu$m). However, the major maximum at $\lambda\cong 3.37$ $\mu$m discussed above, is present no matter how small is the virion and, importantly, it gives almost the same relative extinction power. Such a feature remarks further the significance of the reported resonance, since the electromagnetic beam concentrated around a single wavelength allows for maximal interplay with an ensemble of virus particles possessing various sizes. Note that mid-infrared frequency range makes a privileged band for biosensing \cite{SvetaPaper} and chemical identification of biomolecules through their vibrational fingerprints, namely, photonic operation at this resonance is experimentally feasible \cite{PlasmonicBiosensing}.

\subsection{Power Spatial Distribution}
\begin{figure}[ht!]
\centering
\subfigure[]{\includegraphics[width=4.1cm]{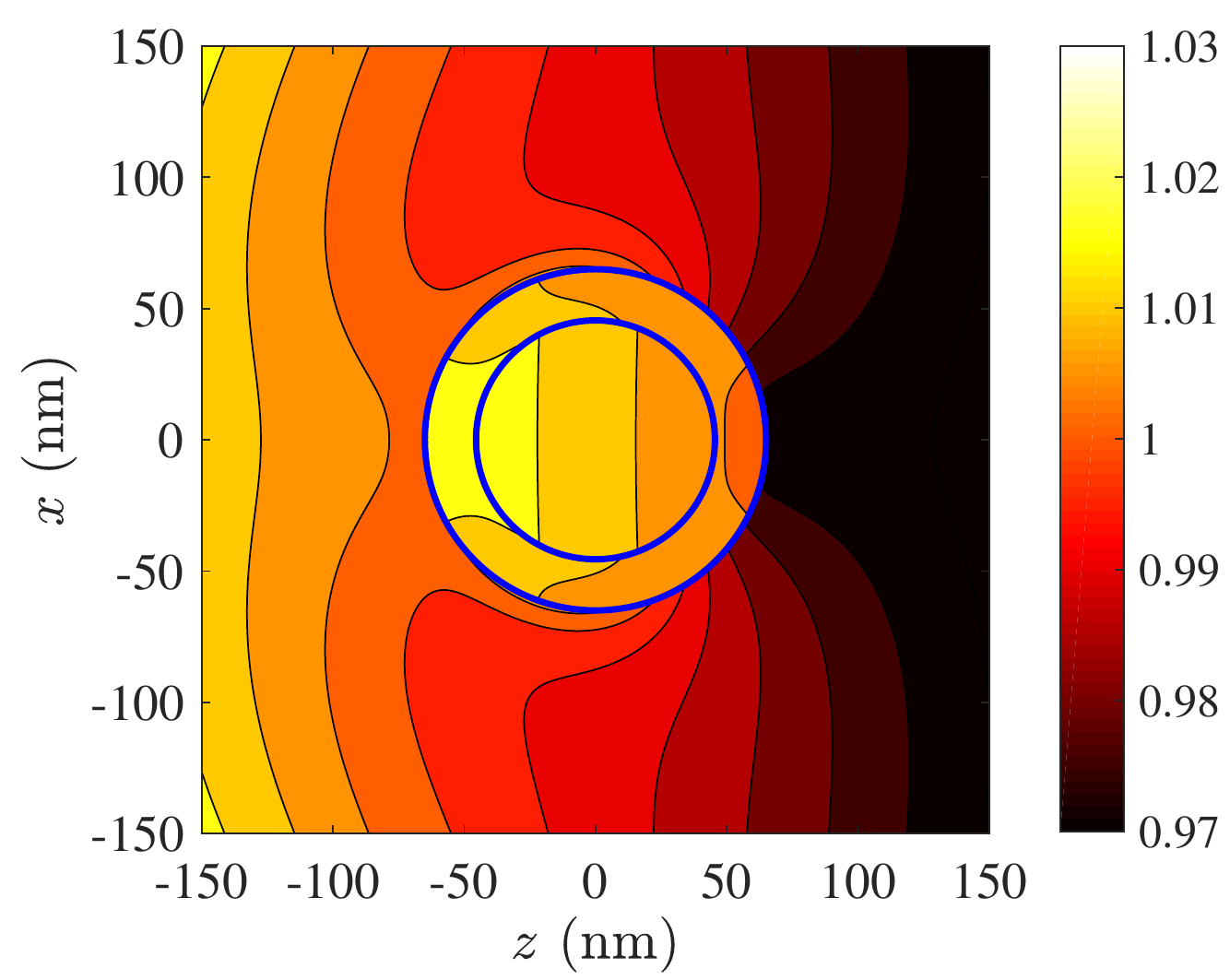}
   \label{fig:Fig4a}}
\subfigure[]{\includegraphics[width=4.1cm]{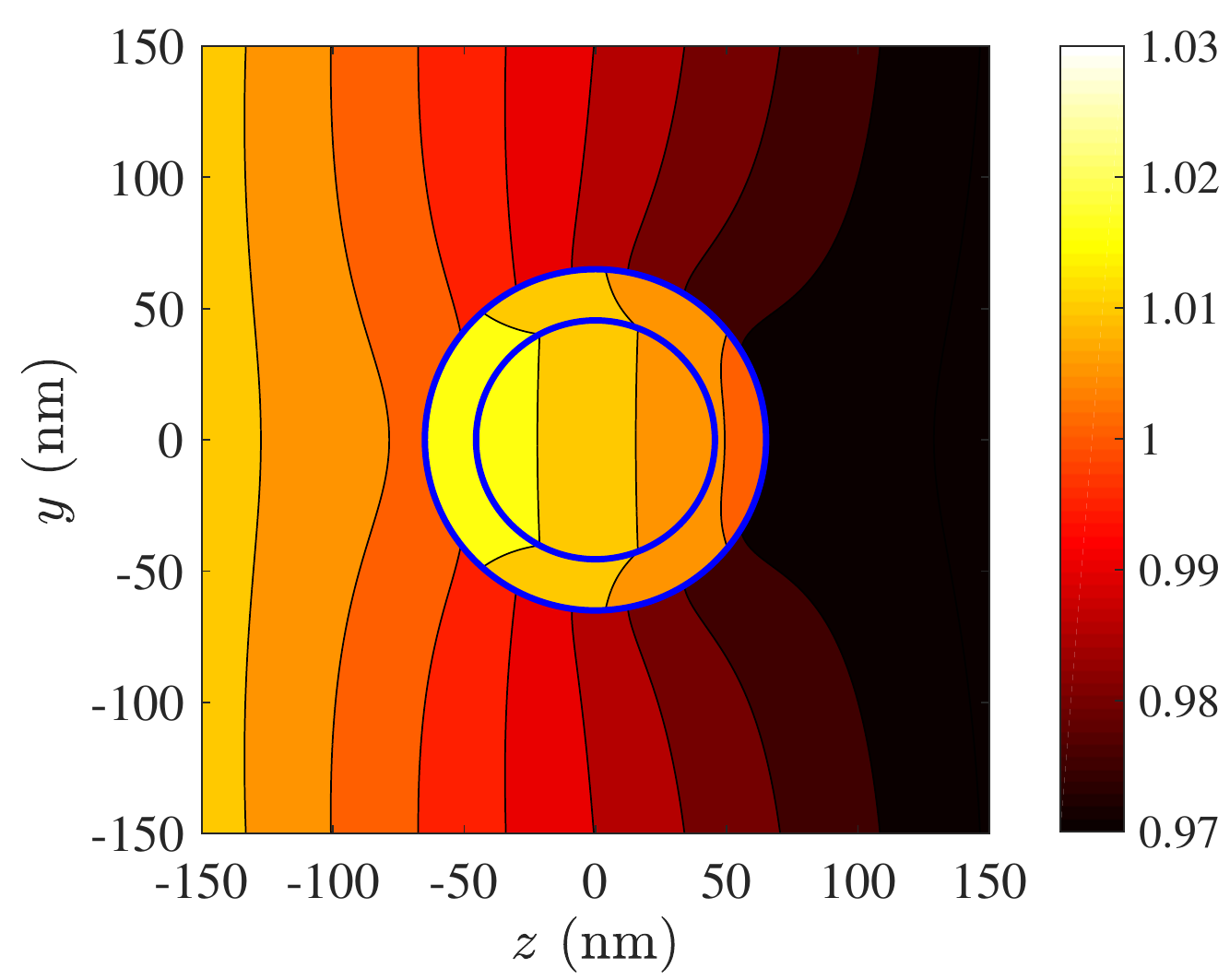}
   \label{fig:Fig4b}}
\subfigure[]{\includegraphics[width=4.1cm]{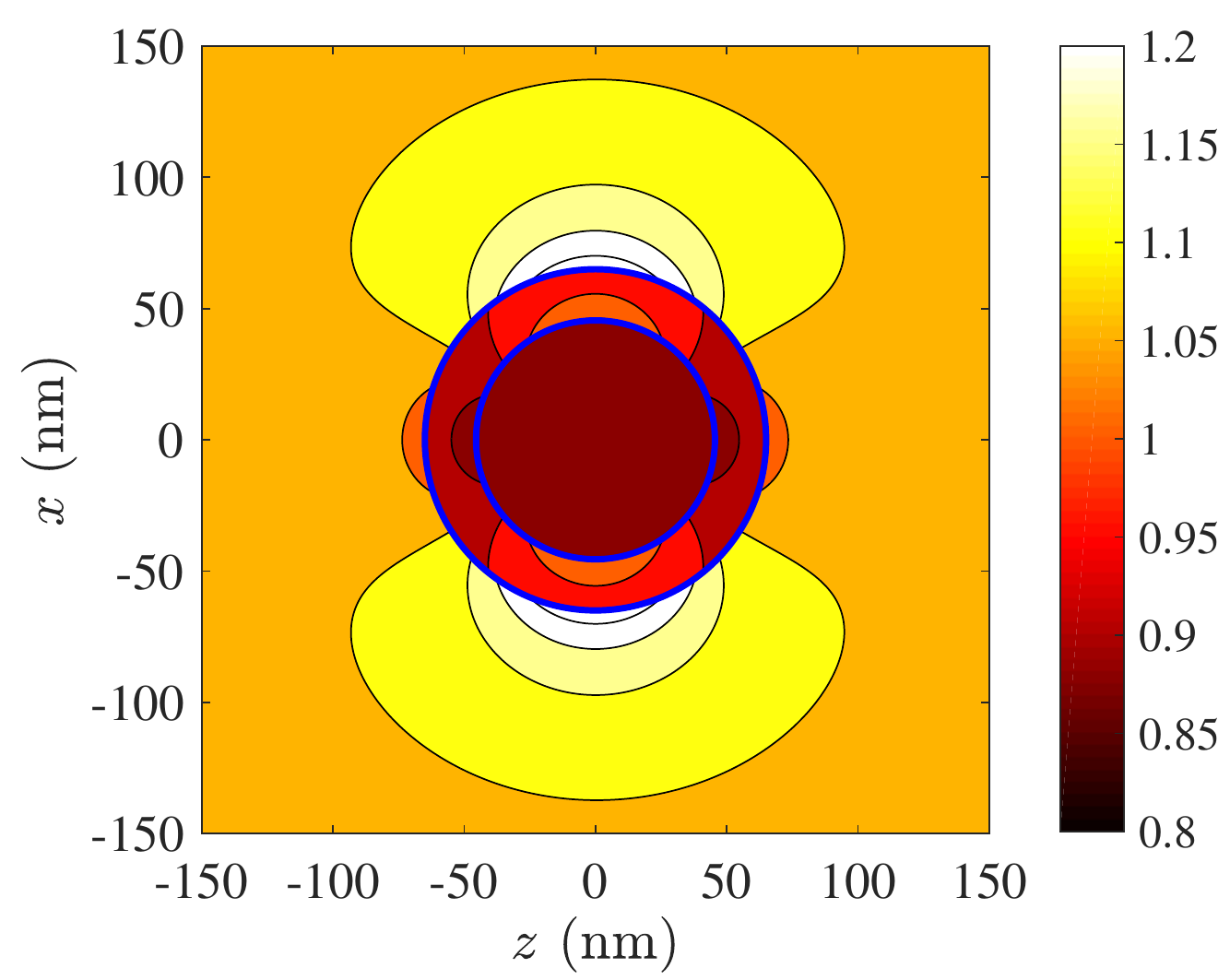}
   \label{fig:Fig4c}}
\subfigure[]{\includegraphics[width=4.1cm]{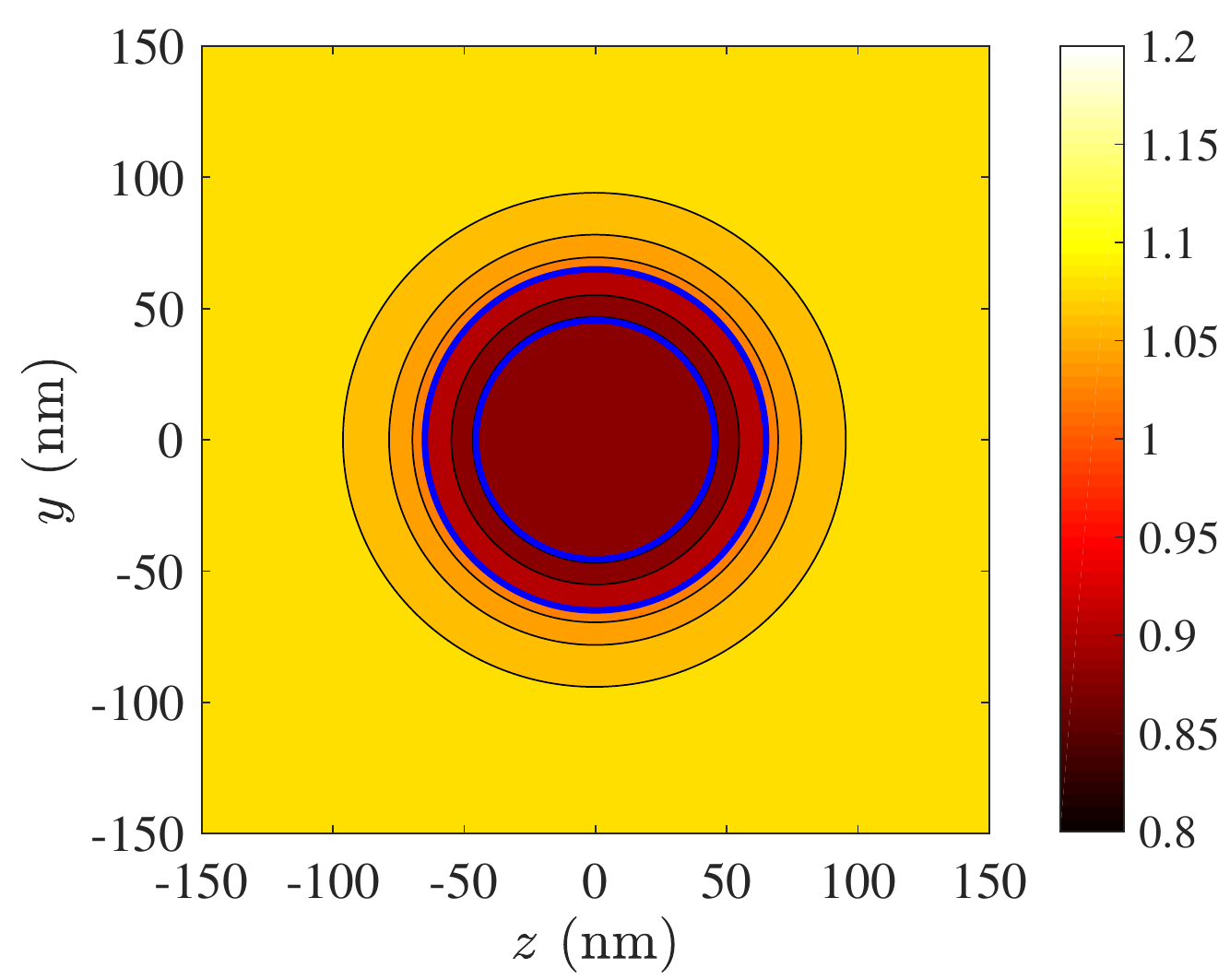}
   \label{fig:Fig4d}}
\caption{Spatial distribution of the total electric field $|\textbf{E}/E_0|^2$ across $zx$ and $zy$ planes when the incoming pulse travels along $+z$ axis for: (a,b) $\lambda=3370$ nm (optimal mid-infrared wavelength) (c,d) $\lambda=2200$ nm (arbitrary smaller wavelength). Plot parameters: $a/b=0.7$, $s=0.5$, $b=65$ nm, human blood background. The blue lines indicate the boundaries between two different media.} 
\label{fig:Figs4}
\end{figure}

After having understood the influence of structural ($a/b$, $s$), textural (several background) and spectral ($\lambda$) parameters on the way that corona-virions interact with the incident beams, it would be meaningful to show the spatial distribution of electromagnetic power inside and outside the core-shell particle for characteristic cases. In Fig. \ref{fig:Fig4a}, we show the relative field quantity $|\textbf{E}/E_0|^2$ across $zx$ plane, when the structure is excited at the optimal mid-infrared wavelength ($\lambda\cong 3.37$ $\mu$m) into human blood host; the represented quantity may be discontinuous as one crosses an interface between two different media, due to the change in texture. By inspection of Fig. \ref{fig:Fig4a}, one directly notices that the values of the electric field magnitude $|\textbf{E}|$ are very close to that of the incident plane wave $E_0$; however, it is natural given the very low permittivity contrast of the virion with its environment. On the contrary, the power concentration at the interior of the spherical volume is counter-intuitive and noteworthy since it clearly demonstrates the substantial interaction of the entire virion with the incoming pulse. As recently reported \cite{OpticalFocusing}, such a physical focus of light is useful to photomedicine while can be directly utilized for thermal damage in biological applications. That significant property is also illustrated when the $zy$ plane is considered (Fig. \ref{fig:Fig4b}) but the distribution is not identical to that of Fig. \ref{fig:Fig4a} due to the vectorial nature of the incident plane wave which is polarized along $x$ axis. 

To show that the bright spots of Figs \ref{fig:Fig4a},\ref{fig:Fig4b} are not easily achieved, in Figs  \ref{fig:Fig4c},\ref{fig:Fig4d} we show the distribution of $|\textbf{E}/E_0|^2$ when the incoming wavelength is arbitrarily picked ($\lambda=2.2$ $\mu$m). Despite the fact that the pulse is faster making the virion look larger and admitting it to develop more complex dynamics, the field into the object is lower than the background level forming a bipolar pattern; as a result, the ability of the primary excitation to engage with the particle is severely diminished compared to the optimal case of Fig. \ref{fig:Fig4a}. Similar conclusions can be drawn by juxtaposing Figs \ref{fig:Fig4c},\ref{fig:Fig4d} referring to $zy$ plane; in the scenario of randomly selected wavelength, the pattern is omni-directional and the power is weak into the scatterer, contrary to the signal focusing exhibited in Fig. \ref{fig:Fig4d}.

\begin{figure}[ht!]
\centering
\subfigure[]{\includegraphics[width=4.1cm]{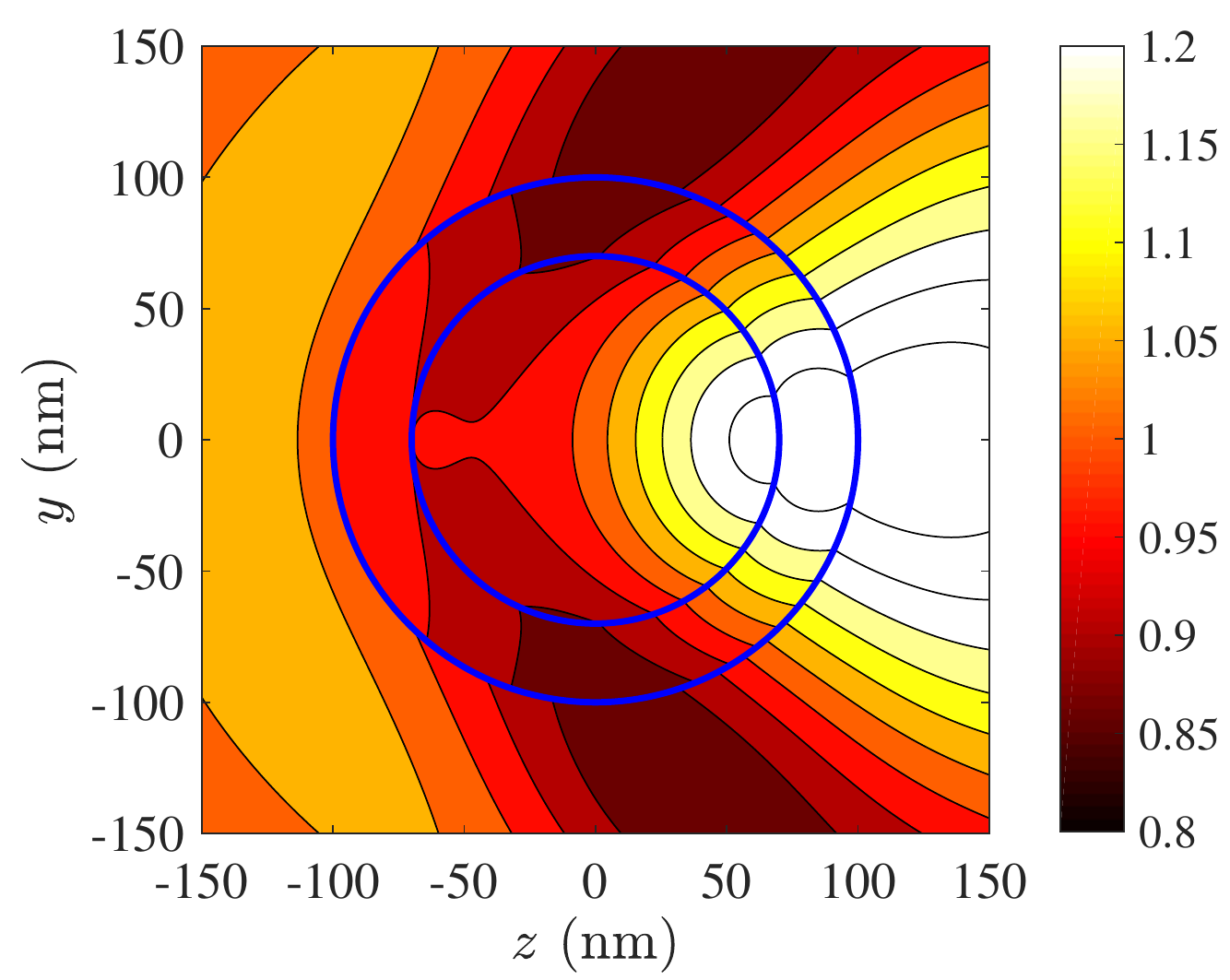}
   \label{fig:Fig5a}}
\subfigure[]{\includegraphics[width=4.1cm]{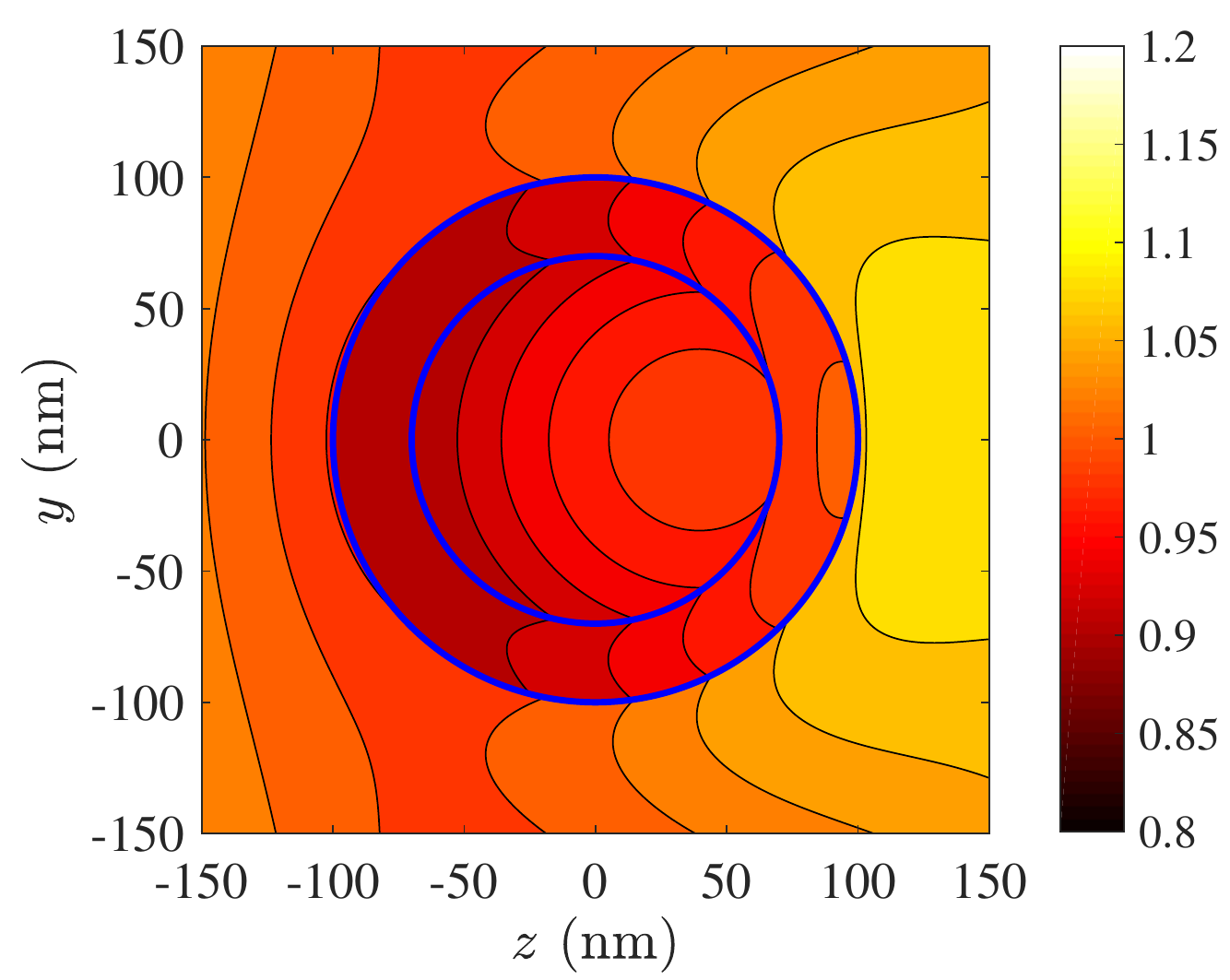}
   \label{fig:Fig5b}}
\caption{Field concentration at shorter wavelengths. Spatial distribution of the total electric field $|\textbf{E}/E_0|^2$ across $zy$ plane when the incoming pulse travels along $+z$ axis for: (a) $\lambda=350$ nm, (b) $\lambda=700$ nm. Plot parameters: $a/b=0.7$, $s=0.5$, $b=100$ nm, human blood background. The blue lines indicate the boundaries between two different media.}
\label{fig:Figs5}
\end{figure}

In Fig. \ref{fig:Figs5}, higher frequencies are examined; more specifically, in Fig. \ref{fig:Fig5a} we show the electric field across $zx$ plane under violet color illumination ($\lambda=350$ nm). One directly notices that the field is stronger at the rear side of the virion but, if an averaging is performed, the relative signal within the volume of the particle is comparable to that of Fig. \ref{fig:Fig4a}, even though the incident wavelength is ten times smaller and much less harmful for the surroundings (healthy cells, tissues, organs) into human body. It is also found that the field increases along $z$ axis into the homogeneous core, contrary to what is happening to Figs \ref{fig:Fig4a},\ref{fig:Fig4b}, where the opposite trend is recorded. In Fig. \ref{fig:Fig5b}, a larger wavelength corresponding to red color ($\lambda=700$ nm) is assumed; we observe a substantially poorer relative power concentration compared to both Figs \ref{fig:Fig4a},\ref{fig:Fig5a} for a much faster pulse than the one operated at the optimal regime.

\begin{figure}[ht!]
\centering
\subfigure[]{\includegraphics[width=4.1cm]{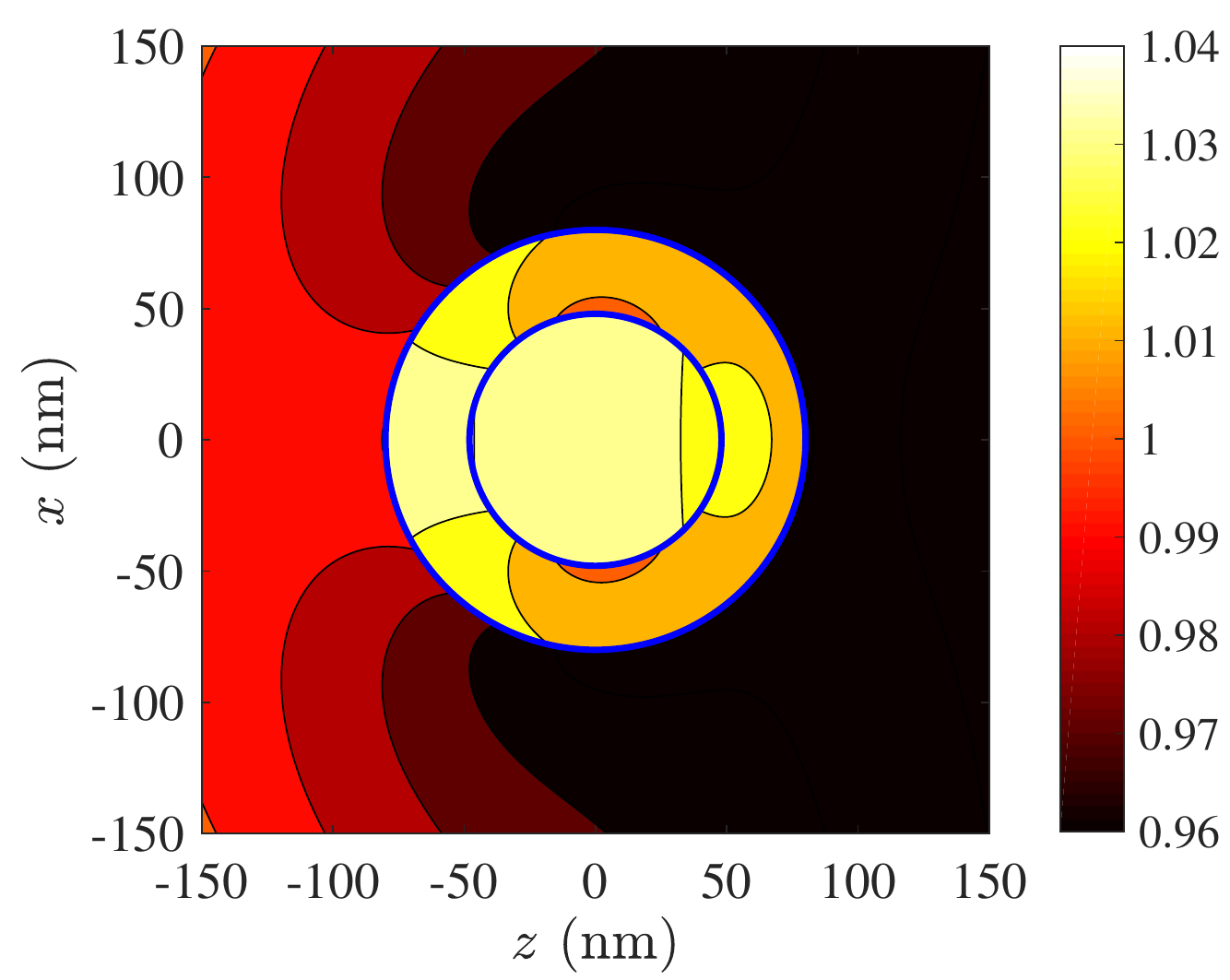}
   \label{fig:Fig6a}}
\subfigure[]{\includegraphics[width=4.1cm]{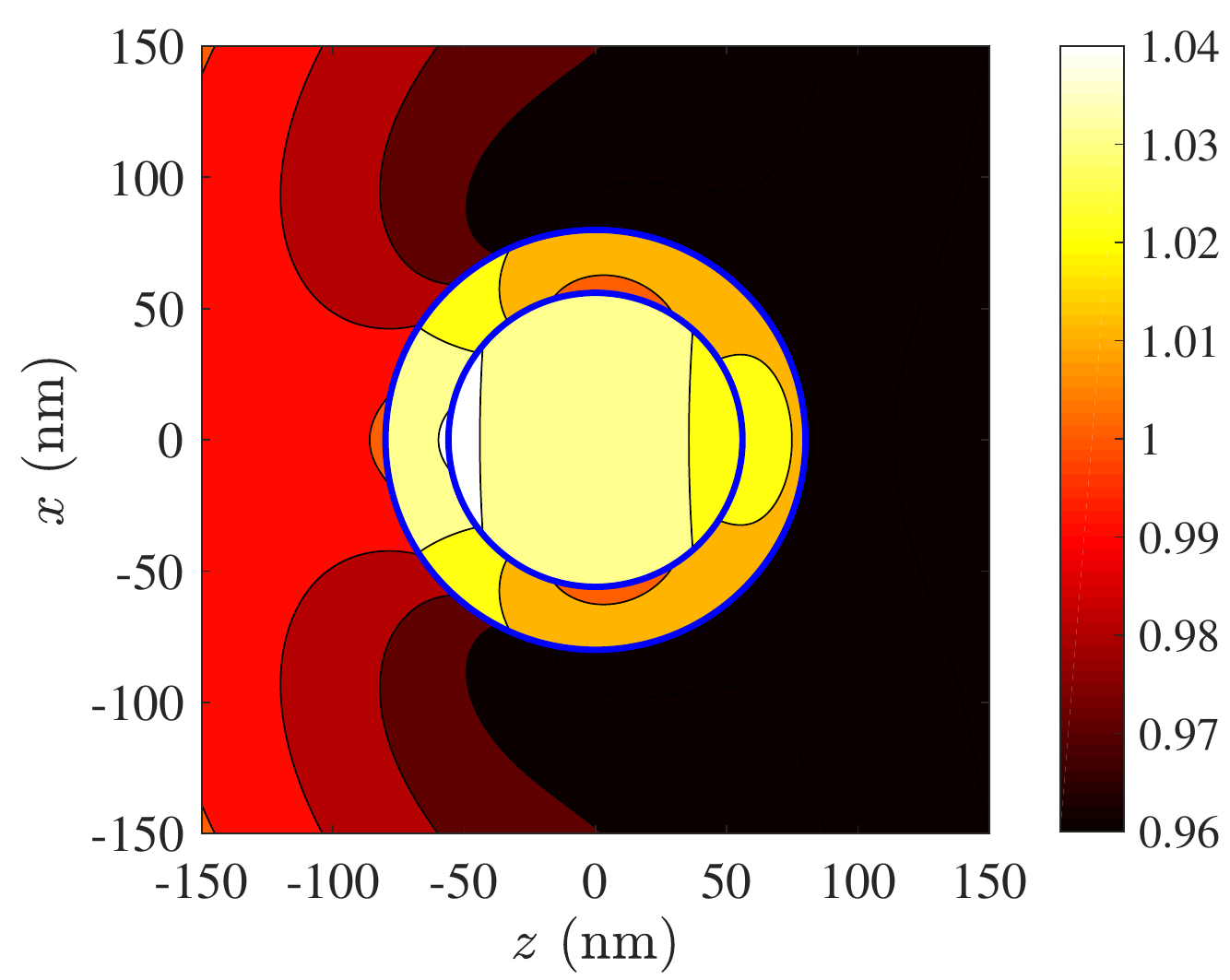}
   \label{fig:Fig6b}}
\subfigure[]{\includegraphics[width=4.1cm]{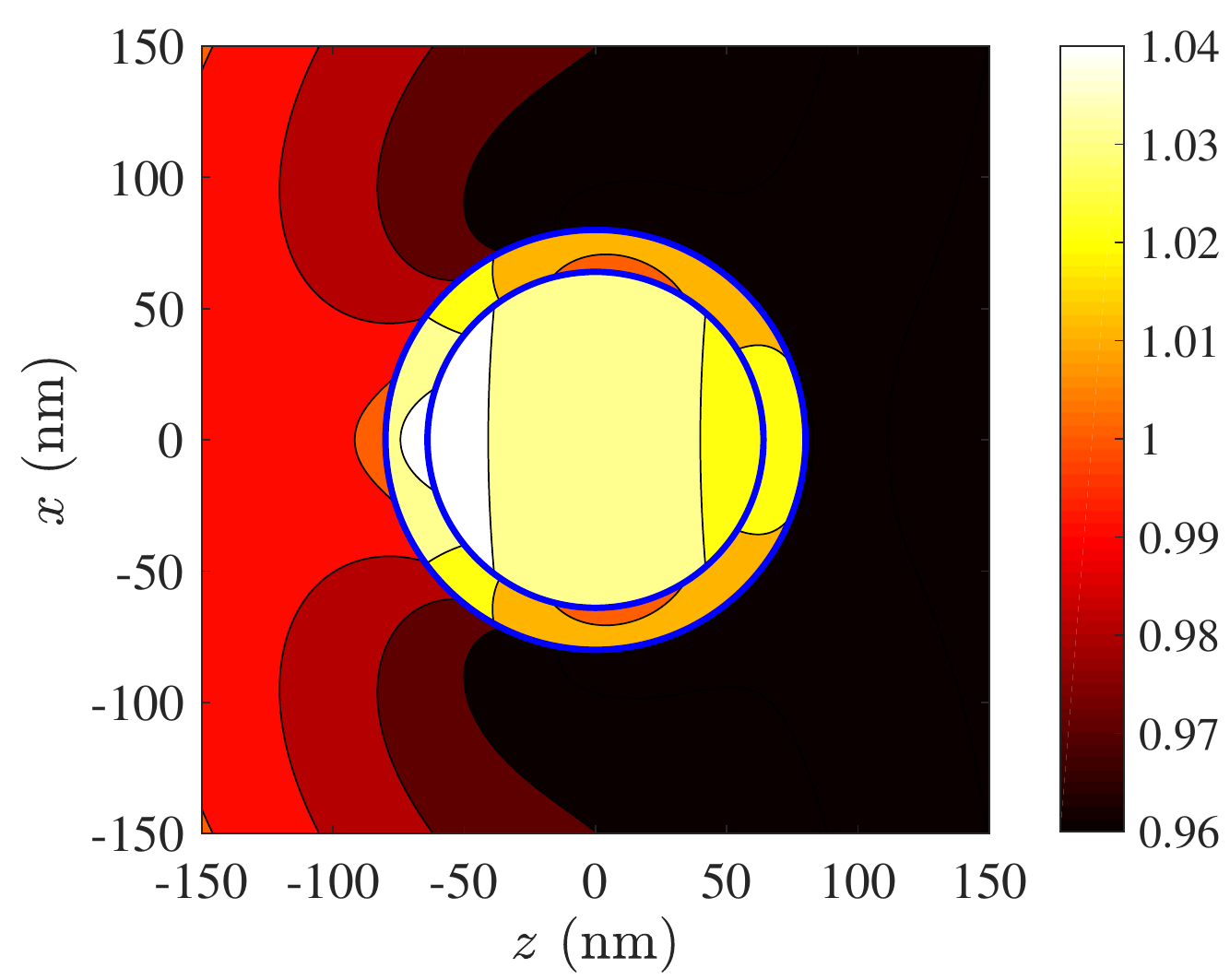}
   \label{fig:Fig6c}}
\subfigure[]{\includegraphics[width=4.1cm]{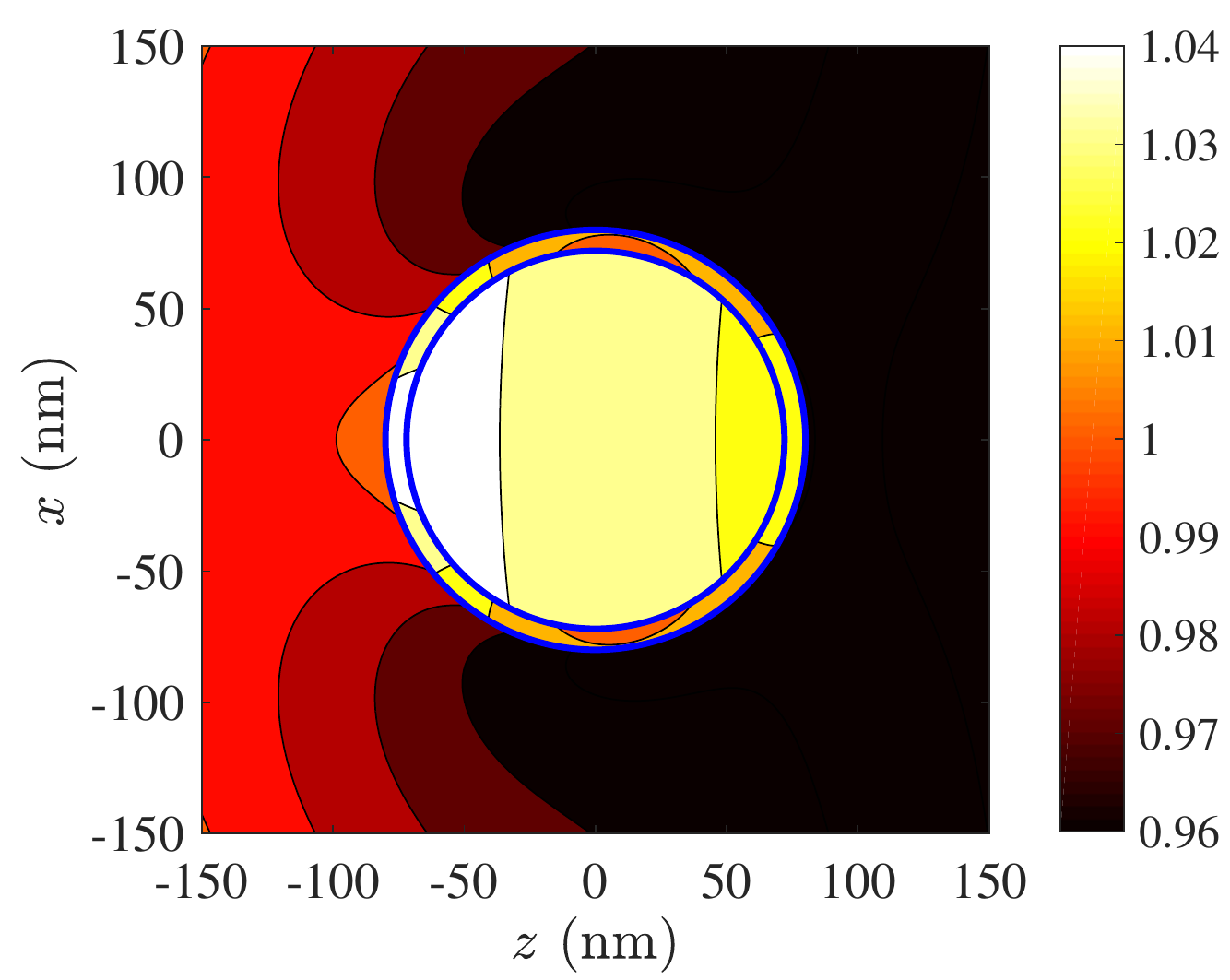}
   \label{fig:Fig6d}}
\caption{Effect of aspect ratio on the field concentration at optimal mid-infrared wavelength. Spatial distribution of the total electric field $|\textbf{E}/E_0|^2$ across $zx$ plane when the incoming pulse travels along $+z$ axis for: (a) $a/b=0.6$, (b) $a/b=0.7$, (c) $a/b=0.8$, (d) $a/b=0.9$. Plot parameters: $s=0.6$, $b=80$ nm. $\lambda=3367$ nm, human blood background.} 
\label{fig:Figs6}
\end{figure}

In Fig. \ref{fig:Figs6}, we examine the effect of the radii ratio $a/b$ on the spatial distribution of the electric field $|\textbf{E}/E_0|^2$, when working close to the reported optimal point ($\lambda\cong 3.37$ $\mu$m). It is apparent that the power accumulation is significant in all the considered cases and gets higher for an increasing $a/b$, as also indicated by Fig. \ref{fig:Fig2a}. In the same way that Fig. \ref{fig:Fig3b} shows insensitivity of the identified effect from the viral size $b$, Fig. \ref{fig:Figs6} designates that the resonance is practically indifferent to the corona thickness $(b-a)$ and thus a universal effectiveness against large families of virions possessing diverse characteristics, is demonstrated. Such an optical trap allows for manipulation and energy exchange with individual viral nanoparticles that can lead to their disintegration or neutralization; similar biomagnifying effects are reported for different applications \cite{SingleCell} like optical imaging and  assembly of bionanomaterials. The experimental potential of the described effect is further underlined by the fact that this type of resonances make typical spectrophotometers to be sensitive in the activity of individual enzymes; thus, it renders the device suitable \cite{OnChipSpectro} for on-chip antibody or antigen detection and chromogenic-based operations such as bacteria detection.

\section{\label{concl} Concluding Remark}
A single corona virion is modeled as a homogeneous spherical core surrounded by a conformal isotropic shell, under illumination from electromagnetic pulses. Rigorous Mie theory is applied to obtain the analytical solution of the formulated boundary value problem and compute the total extinction power of the particle. This quantity indicates how successfully the considered cells interacts with the incoming fields and is found to be maximal at a specific mid-infrared resonance that is independent from the structural characteristics or the background host. A substantial power exchange between impinging beams and the object is necessary for a series of actions dealing with the virion from thermal damage and dissolution to neutralization and isolation; therefore, the reported findings may pave the way to more efficient radiation treatments against SARS-CoV-2.

The proposed model can be refined to include anisotropic multilayers both at the core for more detailed description of the engulfed genome and at the shell to take into account the radial distribution of the protein spikes. Such a process will involve complex-ordered Bessel functions calling for careful computation \cite{WatsonTransform}; similarly, alternative excitation beams or time-restricted causal pulses instead of plane waves can be taken into account by properly evaluating complex Fourier integrals \cite{GaussianPulse}. Importantly, an interesting follow-up of our approach would be to regard clusters of cells and investigate their collective dynamics by implementing suitable transforms  for the summation \cite{Summation1, Summation2} of the responses from randomly or deterministically placed cells \cite{MontiPaper}. In this way, the work at hand can be considered as the first step towards the successful modeling of corona-virions and the derivation of analytical formulas for the exchanged power which will simplify significantly the subsequent optimizations and make easier the designation of optimal operation regimes.

\begin{acknowledgments}
This work was partially supported by Nazarbayev University Grant No. 090118FD5349 entitled: ``Super transmitters, radiators and lenses via photonic synthetic matter''. Funding from MES RK state-targeted program BR05236454 is also acknowledged.

The data that supports the findings of this study are available within the article.
\end{acknowledgments}

\end{document}